\documentclass[twocolumn]{aastex631}

\hypersetup{linkcolor=blue,citecolor=blue,filecolor=cyan,urlcolor=blue}

\newcommand{\ergps}{erg s$^{-1}$}

\newcommand{\mpcc}{$m_{\rm p}$ cm$^{-3}$}
\newcommand{\kmps}{km s$^{-1}$}
\newcommand{\sgrA}{Sgr A$^*$}
\newcommand{\Ledd}{$L_{\rm Edd}$}

\newcommand{\msun}{M$_\odot$}
\newcommand{\oviii}{\textsc{O}\textsc{viii}}
\newcommand{\ovii}{\textsc{O}\textsc{vii} }
\newcommand{\mefeb}{E_{\rm FEBs}}
\newcommand{\efeb}{$\mefeb$}
\newcommand{\etal}{\textit{et. al.}}
\newcommand{\mpy}{\msun yr$^{-1}$}

\shorttitle{Tilted Jets \& Fermi Bubbles}
\shortauthors{Sarkar et al.}
\begin{document}

\title{Misaligned jets from Sgr A$^*$ and the origin of Fermi/eROSITA bubbles}
\correspondingauthor{Kartick C. Sarkar}
\email{kcsarkar@tauex.tau.ac.il, kartick.c.sarkar100@gmail.com}

\author{Kartick C. Sarkar}
\affiliation{School of physics and Astronomy, Tel Aviv University, Tel Aviv, Israel}
\affiliation{Racah Institute of Physics, The Hebrew University of Jerusalem, Israel}

\author{Santanu Mondal}
\affiliation{Indian Institute of Astrophysics, 2nd Block, Koramangala, Bangalore 560034, India}

\author{Prateek Sharma}
\affiliation{Department of Physics and Joint Astronomy Program, Indian Institute of Science, Bangalore 560012, India}

\author{Tsvi Piran}
\affiliation{Racah Institute of Physics, The Hebrew University of Jerusalem, Israel}

\begin{abstract}
One of the leading explanations for the origin of Fermi Bubbles is a past jet activity in the Galactic center supermassive black hole Sgr A$^*$. The claimed jets are often assumed to be perpendicular to the Galactic plane. Motivated by the orientation of pc-scale nuclear stellar disk and gas streams, and a low inclination of the accretion disk around Sgr A$^*$ inferred by the Event Horizon Telescope, we perform hydrodynamical simulations of nuclear jets significantly tilted relative to the Galactic rotation axis. 
The observed axisymmetry and hemisymmetry (north-south symmetry) of Fermi/eROSITA bubbles (FEBs) due to quasi-steady jets in Sgr A$^*$ can be produced if the jet had a super-Eddington power ($\gtrsim 5\times 10^{44}$ erg s$^{-1}$) for a short time (jet active period $\lesssim 6$ kyr) for a reasonable jet opening angle ($\lesssim 10^\circ$). Such powerful explosions are, however,  incompatible with the observed \ion{O}{8}/\ion{O}{7} line ratio towards the bubbles, even after considering electron-proton temperature non-equilibrium. We argue that the only remaining options for producing FEBs are i) a low-luminosity ($\approx 10^{40.5-41}$ erg s$^{-1}$) magnetically dominated jet or accretion wind from the Sgr A$^*$, 
and ii) a SNe or TDE driven wind of a similar luminosity from the Galactic center.
\end{abstract}

\keywords{ISM: jets and outflows -- Galaxy: centre -- Galaxy: halo}

%
%
%
%
\section{Introduction} 
\label{sec:intro}
The decade-old discovery of two giant gamma-ray bubbles toward the Galactic Center (GC) called the Fermi Bubbles \citep[FBs;][]{SuEtAl10, Ackermann14, Selig2015} 
spurred a discussion of their origin. The recent discovery of X-ray structures in the southern Galactic hemisphere by e-Rosita \citep{Predehletal2020Natur}, which appear to be the counterparts of the known X-ray features in the northern hemisphere, has reignited the quest for the origin of the Fermi/eROSITA bubbles (FEBs). Leading models are (i) the star-formation wind-driven scenario where overlapping supernovae and massive stellar winds produce a biconical outflow perpendicular to the Galactic disk \citep{Lacki14, Crocker2015, Sarkaretal15b, Sarkar2017, Sarkar19} and (ii) the central supermassive black hole (SMBH; Mass $M_{\rm bh} \approx 4\times 10^6$ \msun) wind/jet-driven scenarios where FEBs are powered by accretion on to the SMBH \citep{Guo12, ZubovasNayakshin12, Mouetal14, Keshetgurwich17, ZhangGuo2020, Mondal2022,Yang2022}. These models have varying degrees of success in reproducing the FEBs in different wavebands.

While the star-formation driven scenarios require a star-formation rate close to the observations \citep{NoguerasLara2020}, the SMBH jet/wind-driven scenarios typically assume an enhanced mechanical power ($\sim 10^{-3}-10^{-1} L_{\rm Edd}$) of the SMBH in the recent past since the current mechanical luminosity (power) of \sgrA is estimated to be $10^{-8}-10^{-6}$ \Ledd \citep{Agol2000, Marrone2006}.   

The strongest support for the SMBH jet-driven scenario is the presence of sub-parsec nuclear stellar disks, sub-parsec gaseous streams, and the large-scale ionization cone in the Galaxy \citep{Paumardetal06, Genzel10, Bland-Hawthornetal19}, indicating a past accretion event in \sgrA about $\sim$ a few Myr ago. The possible jet activity from this event could have created FEBs. However, in current jet-driven FB simulations, the jets are injected perpendicular to the Galactic plane (parallel to the Galactic rotation axis), 
while observations of the sub-parsec stellar disks/streams find that their rotation vectors are inclined at an angle of $24^\circ-45^\circ$ from the Galactic rotation axis \citep[see][for a summary of the sub-parsec structures]{Bartko2009, Genzel10}. 
Recent EHT mm images of  \sgrA indicate that the accretion disk angular momentum at a few gravitational radii is directed at an angle of $\gtrsim 60^\circ$ away from the Galactic rotation axis \citep{Akiyama2022}. It is, therefore, more reasonable to assume that a possible nuclear jet launched by \sgrA would be significantly misaligned relative to the Galactic rotation axis.
Recently, galaxy formation simulations have been able to resolve gaseous dynamics around a SMBH as the gas is accreted from kpc scale to the sub-pc scale. The simulations find that the sub-pc accretion disk around the SMBH in the AGN phase is tilted by $0-60^\circ$ with a time-averaged mean at $\sim 35^\circ$ with respect to the kpc-scale gaseous disk 
\citep{Angles-Alcazar2021}.
All the evidence, therefore, suggests that a plausible jet from the SMBH is most likely to be directed significantly away from the Galactic rotation axis. 

The observable features produced by a jet in our Galaxy would largely depend on the interaction of the jet and the ambient medium \textit{viz.}, the interstellar and circumgalactic medium (ISM/CGM), and not on just on its original direction. 
However, if the jet-ambient interaction is strong enough, such that the jet is choked before it emerges from the high-density region of the interstellar and circumgalactic medium then all the
jet kinetic energy is converted into the thermal energy of the expanding cocoon. We call this process `jet-dissipation'. 
This happens typically when the jet engine stops before the ejecta emerges from the dense surrounding region, but other possibilities like kink instability of a weak magnetic jet could also lead to dissipation \citep{Tchekhovskoy2016}.
For a dissipated jet, the shape of the cocoon will be determined not by the jet direction but by the density gradient of the ambient medium, which is naturally perpendicular to the Galaxy. 
Since the observed FEBs are axisymmetric (East-West symmetry) and hemisymmetric (North-South symmetry) to a very high degree at the Solar vantage point, they must also be intrinsically symmetric.
Therefore, given that the expected direction of the jet is not perpendicular to the Galactic disk, it is reasonable to conclude that  
dissipation is essential to produce  the observed FEBs.\footnote{In quasi-spherical flows, such as the star formation-driven winds or the accretion winds, the energy injection is inherently isotropic and hence the bubble/cocoon follows the ambient density gradient.} 

The power and the active duration of a jet from the SMBH can vary by many orders of magnitude depending on the available gas and its accretion rate onto the SMBH. 
Given the observed total energy of the FEB, $\mefeb \approx 10^{56}$ erg \citep{Predehletal2020Natur},  the jet duration ($t_{\rm inj}$) and its power ($L_j$) must satisfy:
\begin{equation}
\mefeb = L_j\:t_{\rm inj} \ .
\label{eq:elt} 
\end{equation} 
Thus, more powerful jets should operate for shorter periods. 

Several authors have argued that 
jet dissipation can happen for low-power jets in a highly clumpy medium \citep{Rosen1999, Mukherjee2016, Mukherjee2018, Mukherjee2020, Tanner2022}. However,
the physical setup of these simulations is different from ours (wide-jets and/or clumpy media). Further, these simulations used a rather low resolution that appears insufficient to resolve jet collimation, and it is possible that the results were influenced by this  
(see Appendix \ref{app-sec:prev-works} for a detailed discussion). 
In this paper, we show that only high-power jets ($L_j \gtrsim L_{\rm edd}$) that are choked (i.e., without active injection; since $t_{\rm inj} = \mefeb/L_j$) before the cocoon leaves the ISM, can be dissipated. However, we argue that such high-power jets are inconsistent with the observed \oviii/\ovii line ratio constraints \citep{MillerBregman2016, Sarkar2017} and, therefore, are ruled out as the origin of the FEBs.

This paper is organized as follows. We present the results from our numerical simulations in section \ref{sec:simulations} and show the necessity of jet-dissipation within the ISM for producing symmetric FEBs. We provide analytical arguments supporting our simulations as well as discuss general requirements for dissipation of non-magnetic jets in section \ref{sec:dissipation}. In section \ref{sec:oviii-ovii-ratio}, we discuss possible implications of short-lived and powerful jets with respect to the \oviii/\ovii constraint. In section \ref{sec:discussion}, we consider the fate of magnetic jets, accretion winds, and star formation-driven wind bubbles. Section \ref{sec:conclusion} summarizes our findings on the origin of the FEBs.

%
%
\begin{figure*}
  \centering
 \includegraphics[width=0.97\textwidth, clip=true, trim={2.0cm 0cm 1.0cm 1.3cm}]{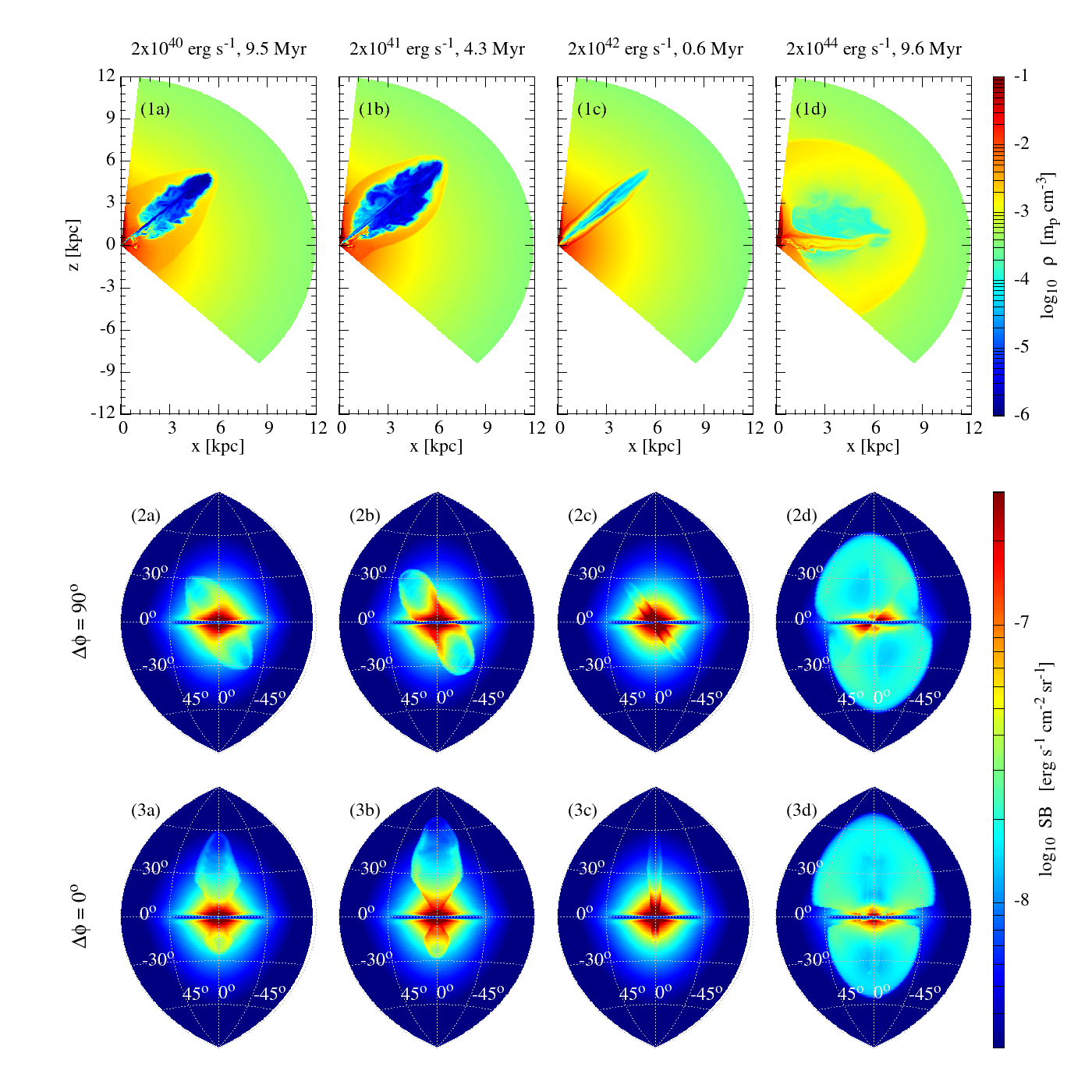}
  \caption{Simulated bubbles for different injected jet power, $L_j =2\times 10^{40}, 2\times 10^{41}$, $2\times 10^{42}$, and $2\times 10^{44}$ \ergps from left to right, at the GC. The jet is switched off once a total jet energy of $10^{56}$ erg has been injected. The top panels show the density contours for different jets. The middle and bottom panels correspond to the projected soft x-ray ($0.5-2.0$ keV) surface brightness maps (assuming \textsc{mekal} plasma model)  for viewing angles $\Delta\phi=90^\circ$ (\textit{out-of-plane} view) and $\Delta\phi=0^\circ$ (\textit{in-plane} view), respectively. 
  The snapshots are taken, respectively, at $t = 9.5$, $4.3$, $0.6$, and $9.6$ Myr. The non-monotonicity in the cocoon age for $2\times 10^{44}$ \ergps case is due to early jet choking (see eq \ref{eq:elt}). The cocoon evolution in this case is different from the active jets (see section \ref{sec:dissipation}).
  The wedge shape at the base of the cocoons is due to the limited range of the simulation box in the $\phi$-direction. We increase the $\phi$-range of our $2\times 10^{44}$ \ergps case to $10^\circ-170^\circ$ for a more realistic x-ray map.}
  \label{fig:Projected-aitoff-three}
\end{figure*}
%

\section{Hydrodynamical simulations} 
\label{sec:simulations}

\subsection{Numerical set up}
\label{subsec:set-up}
We perform 3D hydrodynamical simulations of jets expanding in a realistic ISM/CGM of the Galaxy using the hydrodynamic code \textsc{pluto} \citep{Mignoneetal07}. The simulations are performed in 3D-spherical coordinates with the Galactic disk lying on the $\phi=0,180^\circ$ plane and the rotation axis pointing along the $\theta,\phi = 90^\circ,90^\circ$ direction. This special coordinate setup avoids possible numerical artifacts at $\theta=0, \pi$ axis (coordinate singularities). Although the jets are supposed to be launched at a sub-AU scale in \sgrA, we cannot resolve the launching region while simultaneously modeling $\sim 10$ kpc FEBs in the same simulation box. Our fiducial simulation box, therefore,
extends i) from $10$ pc to $12$ kpc radially with a uniform $2$ pc resolution till $300$ pc and 360 logarithmic grids outside, ii) from $5^\circ$ to $135^\circ$ in $\theta$-direction with $0.5^\circ$ resolution, and iii) from $50^\circ$ to $130^\circ$ in $\phi$-direction with $0.45^\circ$ resolution. We ensure that this resolution is sufficient to resolve the jet dynamics and produce numerically converged bubbles (see Appendix \ref{app-sec:grid-structure} for details on the grid structure). 

The initial density distribution consists of a rotating ISM disk (central density $=1$ \mpcc, and turbulent velocity of $24$ \kmps) and an isothermal CGM (central density $=0.019$ \mpcc and $T = 2\times 10^6$ K). The initial gaseous distribution is in steady-state equilibrium confined by the background gravity of the dark matter, stellar disk, and the stellar bulge. A detailed description of the setup is given in \cite{Sarkaretal15a,Sarkar2017}. We do not include radiative cooling in the simulations since the dynamical time of the shock in ISM or CGM is shorter than the cooling time in the medium. 

As an example of a tilted jet motivated from observational constraints, the hydrodynamical jet is launched at an angle of $45^\circ$ 
from the Galactic pole and has a half opening angle,  $\theta_0 = 10^\circ$. The jet axis lies in $\theta,\phi=45^\circ,90^\circ$ direction.
The jets are launched by continuously adding mass and keeping the velocity of the fluid at $v_j = 0.1 c$ within the solid angle of the jet and in a region $r \leq 30$ pc from the center. The mass addition rate equals the total mass outflow rate ($2 L_j/v_j^2$) of the jet. 

\subsection{Cocoon dynamics}
\label{subsec:cocoon-dynamics}
We perform four simulations with jet power of $2\times 10^{40}$ (low-power; LP), $2\times 10^{41}$ (medium power; MP), $2\times 10^{42}$ (high power; HP), and $2\times 10^{44}$ \ergps (Eddington power; EP), the results for which are shown in figure \ref{fig:Projected-aitoff-three}. Different columns in the figure represent different luminosities. The snapshots are shown at times when the Galactic latitude, $b$, of the outer edge of the shock reaches $\sim 60-70^\circ$ in the Northern hemisphere, corresponding to the size of the eROSITA bubbles.
The density slices for the LP, MP, and HP cases show that the jet clearly `remembers' its injection direction and it produces cocoons that are significantly tilted from the Galactic pole. Although the LP and MP cocoons are  wider than the HP case, they are still significantly tilted as the jets in these cases have not dissipated. 
This is in contrast to the results presented by several authors \citep[c.f.,][]{Mukherjee2018, Tanner2022}, where the low-power jets show dissipation. Based on several simulations of a comparable resolution and a similar jet-injection method as in those papers, we also find dissipation for low-power jets. We stress that the dissipation seen in these Cartesian box simulations could be  
due to the inability to numerically resolve the re-collimation process at the jet base (see appendix \ref{app-sec:prev-works}) 
For us, dissipation happens only for the EP jets where the jet 
engine is turned off and the jet is choked ($t_{\rm inj} \approx 16$ kyr) before it can break out of the ISM (similar to the cocoons dynamics in gamma-ray bursts \citealt{Pais2022}).
Once choked, the dynamics of the cocoons that form in such a short duration burst is very similar to a Sedov-Taylor blast wave, and therefore, it follows the density gradient of the ambient medium (in this case, the ISM/CGM).

\subsection{Projection maps}
\label{subsec:projection-maps}
The tilted cocoons will never produce symmetric FEBs, irrespective of the emission mechanism for $\gamma$-rays. This means that dissipation of the jets and the subsequent vertical rise of the cocoons along the Galactic rotation axis is essential for jet-driven FEBs. 
Projection effects profoundly affect the observational appearance of FEBs. Given the uncertainty on inclination and the position angle of the claimed jet \citep{Genzel10} (although recent EHT images of \sgrA suggest an almost face-on jet pointing towards us), our vantage point will change the appearance of the FEBs. To investigate the influence of such projections, we calculate the soft x-ray ($0.5-2.0$ keV) surface brightness maps from the local density, temperature, and metallicity ($=0.5$ Z$_\odot$) of the gas assuming \textsc{mekal} plasma model \citep{Mewe1985, Mewe1986, Liedahl1995}.
We produce the X-ray surface brightness maps for our simulations from the solar vantage point using the projection software, \textsc{pass} \footnote{Freely available at \url{https://gitlab.com/kartickchsarkar/PASS-EOV}.} \citep{Sarkar2017}. The projection maps also consider the effect of an extended hydrostatic CGM till $100$ kpc that is not included within the simulation box.

The middle and bottom rows of figure \ref{fig:Projected-aitoff-three} show the x-ray surface brightness maps in Aitoff projection from two different Solar vantage points: one, perpendicular to the plane of the jet at $r,\theta, \phi = 8.5$ kpc, $90^\circ, 0^\circ$ (\textit{out-of-plane} projection; middle panel); and second, in the plane of the jet at $r,\theta = 8.5$ kpc, $0^\circ$ (\textit{in-plane} projection: lowest panel), to show the range of possible projection effects. The tilt in the cocoons is  visible in the  \textit{out-of-plane} projection for all  cases, except in EP, for which the tilt is not very apparent, and the cocoon seems to be roughly symmetric \footnote{We also performed a simulation for $2\times 10^{43}$ \ergps for which the cocoons show a significant asymmetry around the Galactic plane and the rotation axis.}.
As expected, the \textit{in-plane} view appears axisymmetric for all  cases. However, since one of the cocoons now approaches the Sun, it is much closer than its counterpart in the other hemisphere. Therefore, the approaching cocoon appears bigger than the receding one, making them non-hemisymmetric, as seen in the lowest panel of figure \ref{fig:Projected-aitoff-three}. 
It is, therefore, clear that the tilted jets either produce axisymmetric cocoons or hemisymmetric cocoons but never both together.
In contrast, when the jets are dissipated due to early jet-choking, the cocoon follows the density gradient which itself follows the axisymmetric and hemisymmetric gravitational potential of the Galaxy \footnote{The slight North-South asymmetry in EP case may be nullified by small density asymmetry in the two hemispheres \citep[c.f.][]{Sarkar19}}.

Based on the above simulations, we conclude that if indeed the FEBs were produced by jet activity at the Galactic center and if the jet was significantly tilted from the Galaxy rotation axis, the jet must have a power $L\gtrsim 2\times 10^{44}$ \ergps or $\gtrsim$ \Ledd. We, however, note that our simulations only consider fixed (although reasonable) values of the jet velocity and opening angle. The condition for jet dissipation will depend on these parameters, and one needs a large suite of simulations to explore the parameter space, which is beyond the scope of the present paper. Instead, we provide theoretical arguments to understand the process of dissipation in the next section and obtain general limits on the jet power.

It is important to note that even a low-power jet that was ejected perpendicular to the Galactic disk and that had not dissipated early enough, would reach the height of the FEBs (i.e. $\sim 10$ kpc) at $\approx 10$ Myr (for LP case) and $\approx 5$ Myr (for MP case), as evident from figure \ref{fig:Projected-aitoff-three}. In such cases, the total injected energy of the cocoon would be $\approx 6\times 10^{54}$ erg (LP case) and $3\times 10^{55}$ (MP case) erg, much less than the energy of the FEBs (\efeb $= 10^{56}$ erg). Additionally, the cocoon would have been elongated and it would not have resembled the observed FEBs.

\begin{figure}
    \centering
    \includegraphics[width=0.45\textwidth, clip=true, trim={4.5cm 2.5cm 4cm 1cm}]{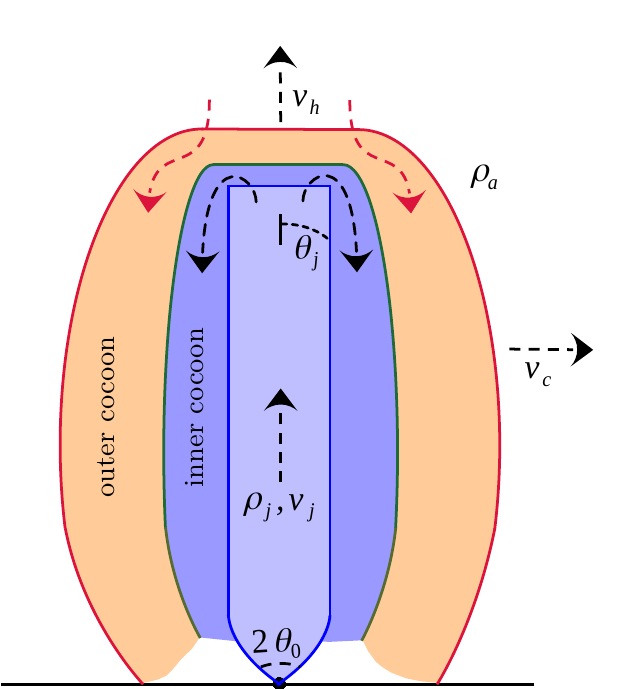}
    \caption{Cartoon of the cocoon geometry. The jet is assumed to be launched with a half opening angle, $\theta_0$, and velocity, $v_j$. The jet density and half opening angle just before the termination shock are $\rho_j$ and $\theta_j$. The cocoon propagates in an ambient medium with density $\rho_a$. 
    }
    \label{fig:cocoon}
\end{figure}

\section{Dissipation of the nuclear jet}
\label{sec:dissipation}
Propagation of jets in an ambient medium and the dynamics of the cocoon is relatively well studied in the literature \citep[e.g.][]{Begelman1989, Marti1997, Matzner2003, Bromberg2011}. Here we reproduce some parts of the calculation that are relevant to the current paper.
 The relativistic jet material interacts with the
ambient medium, creating  a strong shock in its front. It can be shown (see later) that the shock at the jet-head moves much slower than the jet material. The jet material is deflected sideways by the shocked ambient material, which creates a reverse shock that converts most of the jet kinetic energy into thermal energy. This shocked jet material forms a cocoon (known as inner cocoon) surrounding the high velocity jet (see figure \ref{fig:cocoon}). The shocked ambient medium on the other hand produces an outer cocoon that forms an outer layer surrounding the inner cocoon. The inner and outer cocoons are separated by a  contact discontinuity.
\footnote{The terminology `cocoon' differs between the Blazar and GRB communities. In the Blazar community, the cocoon only means the inner cocoon, whereas, the GRB community refers both to the inner and outer cocoons together as the 'cocoon'. Here we follow the GRB community convention.}.

For a jet with a power $L_j$, half opening angle $\theta_j$ (at the jet head), and a jet velocity $v_j$, we can write  the head velocity of the cocoon as \citep{Marti1997, Matzner2003, Bromberg2011},  
\begin{equation}
    \beta_h = \beta_j\: \frac{\Gamma_j \sqrt{\chi}}{1+\Gamma_j\sqrt{\chi}} \approx \beta_j\:\Gamma_j \sqrt{\chi},
    \label{eq:beta_h}
\end{equation}
where $\beta = v/c$, $\chi = \rho_j/\rho_a$ (for cold jets), and $\rho_a$ is the ambient density (in this case, ISM/CGM). The suffix $h$ and $j$ represent the head and the jet. The value of $\chi$ (jet density contrast relative to the ambient medium) can be obtained by considering that the jet luminosity (including the jet in the opposite direction) is given by
\begin{eqnarray}
   L_j  &=& \pi\theta_j^2 z_h^2 \:\rho_j \beta_j^3 c^3 \quad \mbox{(NR)} \nonumber \\
    &=& 2\pi\theta_j^2 z_h^2 \:\rho_j c^3\: \Gamma_j^2  \quad \mbox{(ER)}\,.
    \label{eq:chi}
\end{eqnarray}
Here, $\Gamma_j$ is the Lorentz factor of the  jet material and $z_h$ is the height of the jet head. The terms NR and ER denote  non-relativistic and extremely relativistic dynamics, respectively. Therefore, Eq. \ref{eq:beta_h} is 
\begin{eqnarray}
    v_h &\approx& 300 \mbox{ km s}^{-1}\, \left(\frac{L_{42}}{n_a}\right)^{1/2} \frac{ \beta_j^{-1/2}}{\theta_{j,10^\circ}\: z_{200 pc}}\quad \mbox{(NR)} \nonumber \\
     &\approx& 210 \mbox{ km s}^{-1}\, \left(\frac{L_{42}}{n_a}\right)^{1/2} \frac{1}{\theta_{j,10^\circ}\: z_{200 pc}} \quad \mbox{(ER)}
     \label{eq:vh}
\end{eqnarray}
where, $L_{42} = L_j/(10^{42} \mbox{\ergps})$, $n_a = \rho_a/(0.6\: m_p)$, $\theta_{j,10^\circ} = \theta_j/10^\circ$, and $z_{200pc} = z_h/200$ pc. Note that the apparent discontinuity in the extremely relativistic and non-relativistic velocities is due to different approximations for the energy in these two regimes since the NR approximation breaks down at $\beta \gtrsim 0.75$, i.e. $\Gamma \gtrsim 1.5$. This discontinuity in the NR and ER case propagates to the other equations as well. It is clear that for typical ISM/CGM conditions, the jet head is non-relativistic i.e., $\Gamma_j \sqrt{\chi} \ll 1$ even if the jet material is moving with relativistic velocities (justifying the second equality in equation \ref{eq:beta_h}).

Note that the half-opening angle, $\theta_j$, at the jet head can be very different from the intrinsic jet opening angle, $\theta_0$, at the base of the jet. \cite{Bromberg2011} showed that
the jets are always collimated from a conical geometry to a cylindrical geometry in their passage through the ISM if (Eq 30 in their paper)
\begin{equation}
    L_j \lesssim 1.5\times 10^{49} \mbox{ \ergps } z_{200pc}^2\: \theta_{j,10^\circ}^{2/3}\: n_a \,.
\end{equation}
This is well above the Eddington luminosity of the \sgrA and we can expect that typical jets from \sgrA will be collimated to a cylindrical shape during their passage through the ISM ($n_a \sim 1$, and $z_h \sim 200$ pc). The opening angle for the collimated jet can be written as \citep[using Eq B3 and B9 of][]{Bromberg2011} 
\begin{equation}
    \theta_j = 0.55^\circ\, L_{42}^{1/6} \theta_{0,10^\circ}^{8/15} z_{200pc}^{-1/3} n_a^{-1/6},
    \label{eq:thetaj}
\end{equation}
where $\theta_{0,10^\circ} = \theta_0/10^\circ$.
The jet-head velocity can now be written as 
\begin{eqnarray}
    \frac{v_h}{\mbox{\kmps}} &\approx& 5500 \left( \frac{L_{42}}{n_a}\right)^{1/3} z_{200pc}^{-2/3} \theta_{0,10^\circ}^{-8/15} \beta_j^{-1/2}  \mbox{  (NR)} \nonumber \\
    &\approx& 3400 \left( \frac{L_{42}}{n_a}\right)^{1/3} z_{200pc}^{-2/3} \theta_{0,10^\circ}^{-8/15}  \mbox{  (ER)} \,.
    \label{eq:vh-expl}
\end{eqnarray}
The sideways expansion velocity, $v_c$, of the cocoon can be estimated from the cocoon pressure, $P_c$ ($= 3/4 \rho_a v_c^2$. Since the total thermal energy of the cocoon is simply the energy injected by the jet at any given time, $P_c = (\gamma-1) L_j t/(\pi R_c^2 z_h)$, where we assumed the cocoon to be a cylinder with radius, $R_c \sim v_c t$ and height, $z_h \sim v_h t$. This allows us to solve for the cocoon expansion velocity as 
\begin{eqnarray}
    \frac{v_c}{\mbox{\kmps}} &\approx& 1420 \left( \frac{L_{42}}{n_a}\right)^{1/3} z_{200pc}^{-2/3} \theta_{0,10^\circ}^{-2/15} \beta_j^{-1/8}  \mbox{  (NR)} \nonumber \\
    &\approx& 1300 \left( \frac{L_{42}}{n_a}\right)^{1/3} z_{200pc}^{-2/3} \theta_{0,10^\circ}^{-2/15}  \mbox{ \quad\quad (ER)} \,.
    \label{eq:vc-expl}
\end{eqnarray}

\subsection{Active jet}
As mentioned, we define jet dissipation as the process in which the jet  energy is deposited in a thermal cocoon and the resulting cocoon follows the large-scale density gradient. In this way, the  outflow forgets the initial direction of the jet. Qualitatively, it happens when the cocoon produced by the jet expands purely due to its  internal pressure rather  than the  ram pressure. 

It can be easily seen from Eq \ref{eq:vh-expl} and \ref{eq:vc-expl} that $v_h/v_c \approx 3.9\: \theta_{0,10^\circ}^{-2/5} \beta_j^{-3/8}$ for non-relativistic jets and $\approx 3\: \theta_{0,10^\circ}^{-2/5}$ for relativistic jets. This means that as long as the jet is active  the cocoon will be always significantly elongated along the jet axis, meaning that the cocoon does not expand purely due to its internal pressure. This is consistent with our simulations of LP, MP, and HP cases which are continuously active (see Fig. \ref{fig:Projected-aitoff-three}).
In the context of the FEBs, we conclude that an active (or more recently switched off) jet in the Galaxy would always maintain its direction and would not produce the observed symmetric FEBs.

\subsection{Choked jet}
Dissipation  will happen if the jet is switched off (choked) before  it  breaks out of the ISM. A choked   jet will deposit its kinetic energy as thermal energy inside the ISM. The subsequent cocoon evolution is then similar to a blast wave which follows the ambient density gradient and can produce axisymmetric and hemisymmetric features around the Galactic plane. The condition for choking and subsequent vertical rise of the cocoon is $t_{\rm inj} \lesssim H_{\rm ISM}/v_h$ where $H_{\rm ISM}$ is the height of the ISM ($\sim 200$ pc).  
As the jet duration depends on the jet power (see Eq. \ref{eq:elt}), there is a critical jet power above which the jet will choke before it can escape from the ISM (using Eq~\ref{eq:vh-expl}):
\begin{eqnarray}
    \frac{L_{\rm choke}}{\mbox{\ergps}} &\approx& 8.6\times 10^{44} E_{56}^{3/2} \theta_{0,10^\circ}^{-4/5} H_{200pc}^{-5/2} n_a^{-1/2} \beta_j^{-3/4} \mbox{  (NR)} \nonumber \\
    &\approx& 5\times 10^{44} E_{56}^{3/2} \theta_{0,10^\circ}^{-4/5} H_{200pc}^{-5/2} n_a^{-1/2} \mbox{  (ER)} \,,
\end{eqnarray}
where, $H_{200pc} = H_{\rm ISM}/200$ pc.
Less powerful jets will escape un-choked. 
Qualitatively, high power jets choke because they are switched off early to produce a given total energy.

\begin{figure}
    \centering
    \includegraphics[width=0.45\textwidth]{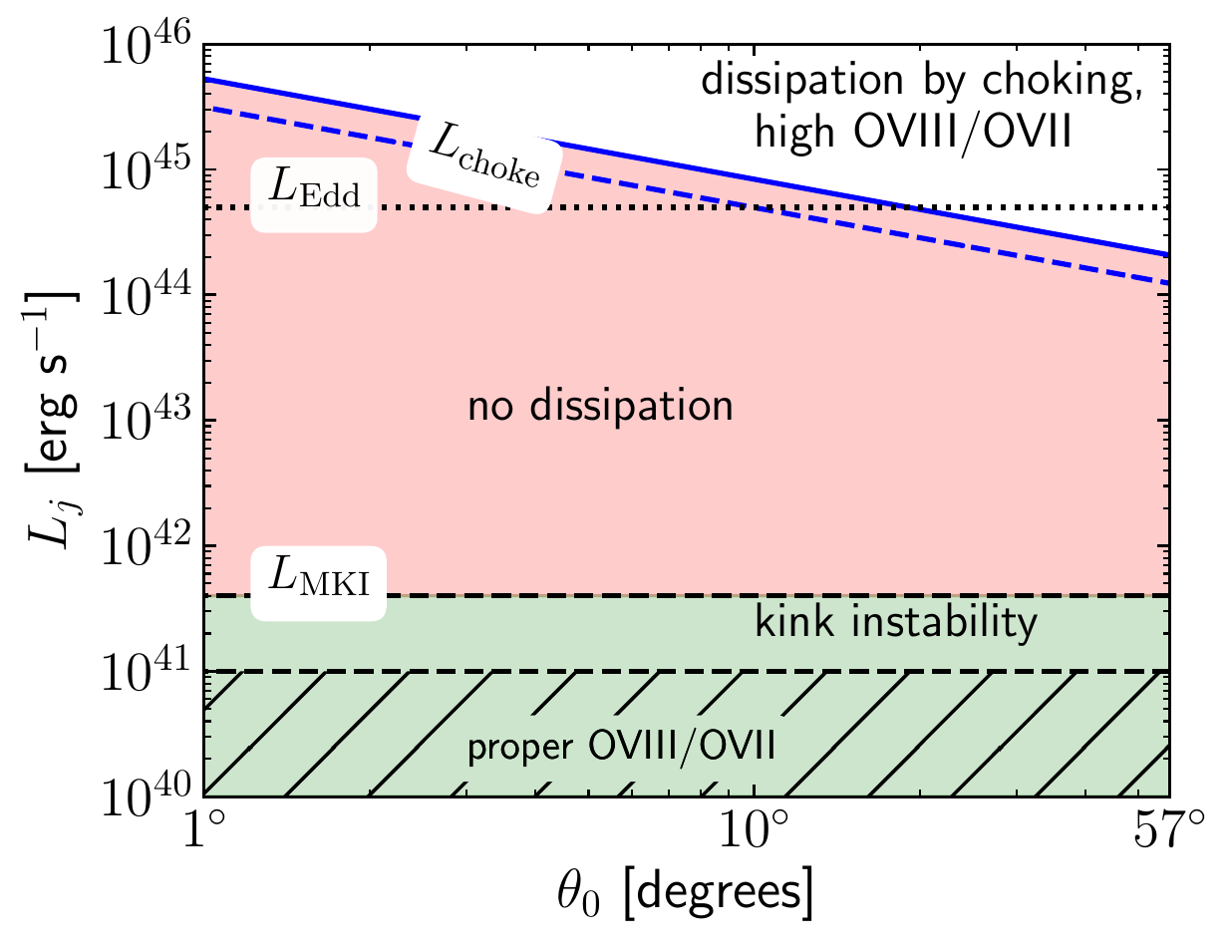}
    \caption{Constraints on the AGN jet luminosity. The red shaded area represents the parameter space for which the jet does not dissipate, and hence it is ruled out.  The blue lines represent the luminosity above which the jet is choked and produces a blast-wave type cocoon, irrespective of the initial jet direction. For a reasonable intrinsic jet opening angle (i.e. $\theta_0 \lesssim 10^\circ$), the jet requires Eddington/super-Eddington luminosity to be choked (since $t_{\rm inj} = \mefeb/L_j$). The blue solid and dashed lines represent the non-relativistic and relativistic limits, respectively. The x-axis extends till $\theta_0 = 1$ radian (57$^0$) since relativistic jets are no longer causally connected above $\theta_0 \gtrsim 1/\Gamma_j$ and behave as a spherical wind. The green region shows where a magnetically dominated jet can be dissipated via kink-instability. The hatched region shows the limit where \oviii/\ovii line ratio is consistent with observations \citep{Sarkar2017}. }
    \label{fig:therm-condition}
\end{figure}

Figure \ref{fig:therm-condition} shows the behaviour of $L_{\rm choke}$ with the intrinsic jet opening angle, $\theta_0$. We see that for reasonable jet opening angles (i.e. $\theta_0 \lesssim 10^\circ$), choking happens at $L_j \gtrsim$ \Ledd. This limit is in excellent agreement with our simulations as shown in figure \ref{fig:Projected-aitoff-three}. Therefore, our analysis so far allows us to conclude that if indeed the axisymmetric and hemisymmetric FEBs were produced by hydrodynamic jet activity in \sgrA, its luminosity must have been above the Eddington limit ($\approx 5\times 10^{44}$ \ergps for MW SMBH) for reasonable values of the jet parameters and the ISM. Such a choked jet is similar to the non-jetted injection in \cite{Mondal2022} or \cite{Yang2022}.

While the simulations provide us a glimpse of the dissipation processes happening for a given set of jet parameters, figure \ref{fig:therm-condition} shows a comprehensive view of the parameter space. The red-shaded region represents where the jet is active for longer period of time and is not dissipated within the ISM. Although the white region above $L_{\rm choke}$ can, in principle, dissipate jets inside ISM, the required jet has to be super-Eddington for a reasonable value of intrinsic jet opening angle ($\lesssim 10^\circ$). In the next section, we show that such super-Eddington jets are not compatible with the x-ray observations.

\begin{figure*}
\centering
\includegraphics[width=1.0\textwidth]{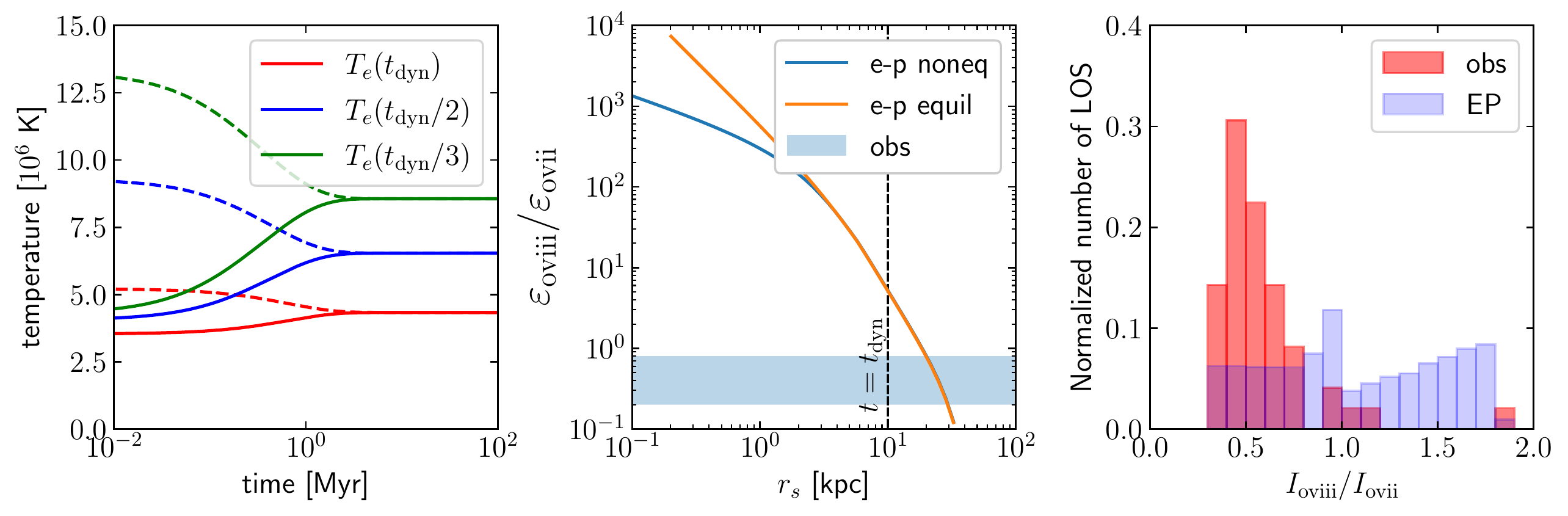}
\caption{\textit{Left}: Evolution of electron ($T_e$; solid lines) and proton ($T_p$; dashed lines) temperatures in the post-shocked gas that is shocked at different times ($t_{\rm dyn}$, $t_{\rm dyn}/2$, $t_{\rm dyn}/3$). Here, $t_{\rm dyn}$ represents the time for the blast wave to reach $R_s = 10$ kpc and the time is measured after this instant. $T_e-T_p$ equilibrium happens almost instantaneously for shocks that are $\gg 1$ Myr old. \textit{Middle}: \oviii/\ovii emissivity ratio corresponding to the electron temperature behind the shock at different stages of its evolution. The orange line shows the ratio if we assume that the $T_e = T_{\rm eq}$ and the blue line shows the values if e-p temperature non-equilibrium is assumed. The blue shaded region shows the observed ratio and the black vertical line shows the expected ratio at $R_s = 10$ kpc. \textit{Right}: histogram of the line ratio in the projected map (for the EP case) and vs observations \citep{MillerBregman2016}. Ratio values $<0.3$ are intentionally removed to avoid the background CGM contribution.}
\label{fig:e-p-noneq}
\end{figure*}

%
%

\section{The \oviii/\ovii ratio}
\label{sec:oviii-ovii-ratio}

One of the main constraints on the dynamics of the FEBs comes from the observed \ion{O}{8}/\ion{O}{7} line ratio towards the bubbles \citep{MillerBregman2016}. Using hydrodynamical simulations, 
\cite{Sarkar2017} showed that the observed line ratio is roughly compatible with power $\approx 10^{40.5-41}$ \ergps~for quasi-spherical flows such as a star-formation driven wind or an accretion wind around \sgrA that is still active. Now turning to highly powerful jets that are choked ($t_j \lesssim 6$ kyr),
the luminosity limit provided by \cite{Sarkar2017} does not apply since the jets are no longer active. Following \cite{Sarkar2017}, we obtain a similar estimate for the \oviii/\ovii line ratio for the choked jets below and show that a super-Eddington, short-duration burst is incompatible with the observed line ratio.

Given that the choked jets have super Eddington power, the events have to be short-lived ($t_{\rm inj}\lesssim E_{\rm FEBs}/L_{\rm Edd} \approx 6$ kyr) to produce the energy of the FEBs. Therefore, we can assume that a choked jet is simply an explosion at the Galactic center and its dynamics can be modeled as a blast wave propagating through the CGM i.e, $r_s = A (E t^2/\rho)^{1/5}$. Here, $A \sim 1$, $E = 10^{56}$ erg is the energy of the blast, $t$ is the time, and $\rho$ is the ambient density. For a power law density profile i.e., $\rho(r) = \rho(r_0)\: (r/r_0)^{-\alpha}$, the above equation can be modified to write down the shock radius and velocity as
\begin{eqnarray}
    r_s &=& A^{5/(5-\alpha)}\: \left(\frac{E}{\rho_0 r_0^\alpha} \right)^{1/(5-\alpha)}\: t^{2/(5-\alpha)} \nonumber \\
    v_s &=& \frac{d}{dt} r_s = \frac{2}{5-\alpha}\: \frac{r_s}{t}
    \label{eq:blast-wave}\,.
\end{eqnarray}
The dynamical time to reach a certain radius $R_s$ is, therefore, $t_{\rm dyn} = A^{-5/2} R_s^{(5-\alpha)/2} \left( \rho_0 r_0^\alpha/E\right)^{1/2}$. For the MW CGM, we assume $\rho_0 = 1.2\times 10^{-2}$ \mpcc, $r_0 = 1$ kpc, and $\alpha = 1.5$ \citep{Miller2015}. We also note that the above blast-wave solution is only valid for an ambient medium with negligible pressure and needs modifications if the ambient medium has significant pressure (which is the case for the CGM). To quantify this, we have run an idealized hydrodynamical simulation of a blast wave in the above power-law density profile using \textsc{pluto} \citep{Mignoneetal07}. We calibrate the value of $A$ against the simulated shock and find that $A = 1.02$ describes the shock well for our purposes here. Therefore, the dynamical time is
\begin{equation}
    t_{\rm dyn} \approx 12 \mbox{ Myr } R_{s,10}^{7/4} E_{56}^{-1/2}
    \label{eq:tdyn}
\end{equation}
where, $E_{56} = E_{\rm FEBs}/10^{56}$ erg.

For strong shocks, it is often the case that the electron temperature, $T_e$,
is not the same as the proton-temperature, $T_p$, or the shock temperature, $T_{\rm eq}$. Such deviation from  non-equilibrium arises
due to a long Coulomb interaction time-scale \citep{Spitzer1956},  
\begin{eqnarray}
    t_{\rm eq} &\sim& 2 \mbox{  Myr }\:\left(\frac{T_e}{4\times 10^6 K}\right)^{3/2}\: \left(\frac{10^{-3}}{n_a}\right) \nonumber \\
    &\sim& 3.3 \mbox{ Myr } \left(\frac{T_e}{4\times 10^6 \mbox{K}}\right)^{3/2} R_{s,10}^{3/2}
    \label{eq:t_eq}
\end{eqnarray}
compared to the dynamical time scale, $t_{\rm dyn}$, of the shock. Here, $R_{s,10} = R_s/10$ kpc and we have used the CGM density profile as mentioned earlier. In such cases, one needs to solve for the electron temperature before comparing it with the observations. 
However, we find that (see appendix \ref{app-sec:ep-noneq} and the left panel of figure \ref{fig:e-p-noneq}) the e-p interaction time-scale for the blast wave, even assuming 
just Coulomb coupling, is short enough (due to a low Mach number of the shock) that the electrons and protons are in equilibrium. Therefore, the shock temperature of the blast wave is a good indicator of the electron temperature for such a blast wave, in contrast to the assumption of $T_e \ll T_p$ needed for \citet{Yang2022} parameters to match the observed \oviii/\ovii line ratio. \footnote{We also solve for the electron temperature for a blast wave with an energy  $E = 10^{57}$ erg (as assumed in \cite{Yang2022}). We find that although $T_e$ lags behind $T_p$ at $t=t_{\rm dyn}$ due to the Coulomb time-scale, in this case, $T_e (t=t_{\rm dyn}) \approx 8\times 10^6$ K. This temperature is much larger than the required value of $\approx 3\times 10^6$ K for explaining the \oviii/\ovii line ratio.}

The middle panel in Fig \ref{fig:e-p-noneq} shows the \ion{O}{8}/\ion{O}{7} line intensity ratio corresponding to the electron temperature at a given radius. The line ratios for given electron temperatures are obtained from \textsc{cloudy}-13.04 \citep{Ferland2013}. The blue line represents the line ratio if the electron temperature is solved till $t=t_{\rm dyn}(R_s)$ starting from the post-shock values at that time, while the orange line simply assumes instantaneous equilibrium. The blue-shaded region represents the observed values \citep{MillerBregman2016}. The figure shows that $T_e$ and $T_p$ are in equilibrium at $r\gtrsim 2$ kpc and that at $R_s = 10$ kpc (roughly the size of the FEBs), the expected \oviii/\ovii ratio is a much higher than the observed range of values. We also confirm this by calculating the projected intensity ratio map of \oviii/\ovii for the EP case (similar to Fig. 5 in \citealt{Sarkar2017}). For a realistic ratio map, we also consider the effect of the extended CGM till $100$ kpc. We show the histogram of the line ratio in the projected map in the right panel of figure \ref{fig:e-p-noneq} and compare it with the observations \citep{MillerBregman2016}. Clearly, such high-energy blast waves are not consistent with the observed \oviii/\ovii and hence do not describe the FEBs.

\section{Discussion} 
\label{sec:discussion}

\subsection{Magnetic Kink-instability}
\label{subsec:kink-instability}
An aspect that we have not discussed so far is the possibility of the magnetic kink-instability \citep[MKI;][]{Bromberg2016,Tchekhovskoy2016}. MKI appears in magnetically dominated jets. It destabilizes the smooth flow of the jet by enhancing instabilities at the jet-cocoon boundary. The time scale to grow MKI  is typically $\sim 10$ Alfven crossing time across the jet width. This indicates a critical jet length after which the laminar jet flow turns into a turbulent one, thus causing jet dissipation. The critical luminosity below which the jet dissipates within a given length scale of the ambient medium (in this case, the ISM) can be written as \citep[Eq 2 of][]{Bromberg2016, Tchekhovskoy2016}
\begin{equation}
    L_{\rm MKI} = 4\times 10^{41} \mbox{ \ergps } n_a\, H_{\rm 200pc}^2 
\end{equation}
for a flat density profile till height $H_{200pc}$. This limit has been shown using the black dashed line in Fig \ref{fig:therm-condition}. This upper limit of the jet luminosity is slightly higher but consistent with the observed \oviii/\ovii line ratio \citep{MillerBregman2016, Sarkar2017} (shown by the black hatched region in figure \ref{fig:therm-condition}). 
Thus a magnetically dominated jet with power $\approx 10^{40.5-41}$ \ergps~can produce the symmetric FEBs that are also consistent with the x-ray observations.

\subsection{Accretion disk wind}
\label{subsec:accretion-wind}
Wind from the accretion disk around the central black hole is
another process that can launch a wide angle ($\sim 30^\circ$) ultra-fast outflow that couples to the ambient medium much more than the jets \citep{King2003, Hopkins2010,King2010, Faucher-Giguere2012, Wagner2013, Jiang2019}.
Such quasi-spherical winds from \sgrA can be another source of energy that can inflate the FEBs and produce the symmetrical features of the FEBs \citep{Yuan2012, Mouetal14, Mouetal15}. \cite{Sarkar2017} showed that the wide-angle winds follow the observed \oviii/\ovii line ratio only if the wind power is $10^{40.5-41}$ \ergps ($\sim 10^{-4}$ \Ledd). Such a wind should sustain for $\sim 30$ Myr (to produce $10^{56}$ erg energy). However, in such a low luminosity AGN (LLAGN) the feedback is mostly dominated by jets rather than by winds \citep{Yuan2012, YuanNarayan2014, Gustini2019}. The jet, on the other hand, would fail to produce the symmetrical features of the FEBs (section \ref{sec:dissipation}, unless it is magnetized and low energy). Therefore, the role of accretion wind as the source of FEBs' energy remains uncertain.

\subsection{Tidal disruption events}
\label{subsec:tdes}
Jets and winds from Tidal disruption events (TDEs) at the central black hole can be a major source of energy for the FEBs. Nuclear star clusters often feed the central black hole with stars for which the stellar orbits usually have a large range of orientations. Thus the resulting jets from each of these events can be directed in a wide area which can result in a quasi-spherical wind if the individual jets do not break out of the ISM. The typical distance traversed by such a jet is (using equation \ref{eq:vh-expl})
\begin{equation}
    z_{\rm tde} \sim 1 \mbox{  pc  } \left( \frac{E_{3e51}}{n_a}\right)^{1/5}\: t_{\rm yr}^{2/5}
\end{equation}
where $E_{3e51} = 3\times 10^{51}$ erg and $t_{\rm yr} = t/1$ yr. Therefore, jets from individual TDEs ($E_{3e51} \sim 1$, $t_{\rm yr} \sim 1$; \citealt{Piran2015}) dissipate their energy well inside the ISM and can act as a thermal energy source for gas dynamics at scales $\gg$ pc. 

It is estimated that a MW type black hole (mass $4\times 10^6$ \msun) has a typical TDE rate of $\sim 10^{-4}$ yr$^{-1}$ \citep{Stone2016}. The total energy produced from such events at the central black hole can be $\sim 3\times 10^{51}$ erg $\times 10^{-4}$ yr$^{-1} \approx 10^{40}$ \ergps. Therefore, either an enhanced rate of TDEs (Piran \etal~\textit{in prep.}) or enhanced energy per TDE would be able to generate enough energy for powering the FEBs \citep{Ko2020}. 

\subsection{Star-formation driven wind}
\label{subsec:sfr-driven}
Supernovae (SN) explosions from the ongoing star-formation at the Galactic center can be another possible source of energy for the FEBs \citep{Crocker2015, Sarkaretal15a}. Since the SNe energy is not directional, dissipation of the energy in the ISM happens naturally. The required star-formation rate to produce the FEBs is $\approx 0.3-0.5$ \mpy over a time-scale of $\approx 30$ Myr \citep{Sarkaretal15a} is well within the observational range of $0.2-0.8$ \mpy within the central $\sim 50$ pc over the last $\sim 30$ Myr \citep{NoguerasLara2020}. 

\section{Conclusions}
\label{sec:conclusion}
We have run 3D-hydrodynamical simulations of jets at the Galactic center that are significantly tilted from the Galaxy rotation axis. The tilted jets are inspired by the observations of sub-pc gaseous/stellar streams and the recent EHT results of the accretion disk around the MW SMBH \citep{Genzel10, Akiyama2022}. Using the simulations and analytical considerations, we show that: 
\begin{itemize}
    \item Jet-dissipation inside the ISM is a necessary condition to produce the observed axisymmetric (symmetry around rotation axis) and hemisymmetric (symmetry around Galactic plane) Fermi/eROSITA bubbles (FEBs).
    \item Jet-dissipation does not happen for non-magnetic jets with power $L_j \lesssim L_{\rm Edd}$. Jets with higher power dissipate their  energy into the ISM via early choking ($t_{\rm inj} \lesssim 6$ kyr) before they break out of the ISM. Such Eddington power jets, however, produce too high \ion{O}{8}/\ion{O}{7} line ratio compared to the observations (even after considering electron-proton temperature non-equilibrium) and are ruled out. 
\end{itemize}
We, therefore, arrive at the conclusion that non-magnetic jets from the \sgrA do not reproduce the morphology as well the x-ray constraints of the FEBs. We speculate that the excess ionization observed in the direction of the Magellanic streams \citep{Bland-Hawthornetal19} might be due to a more recent AGN event ($\sim$ a few Myr ago) that also produced the sub-pc stellar structures and the $\sim 100$ pc radio bubbles in \sgrA \citep{Pontietal2019} but did not produce the FEBs. It is possible that the above $\sim$ Myr jet event is one of many such events that powered the FEBs, but it may not necessarily be the case.

Based on the necessity of jet-dissipation as presented in this paper, we can limit the remaining options for the origin of the FEBs as follows.
\begin{itemize}
    \item \textit{Star-formation driven}: Wind driven by supernovae from the star-forming region at the Galactic center naturally produces symmetric cocoons since supernovae explosions do not have any preferred direction \citep[see][]{Sarkaretal15b, Sarkar2017, Sarkar19}. The required star-formation rate ($\sim 0.5$ \msun yr$^{-1}$) in these models is within the observational limits \citep{NoguerasLara2020}.
    \item \textit{Weak magnetic jets or accretion winds}: If nuclear jets are indeed magnetically dominated, then jet dissipation by kink instability is a possibility. Winds from the past accretion disk naturally satisfy the condition for dissipation. The required magnetic-jet/accretion-wind power (i.e. $\approx 10^{40.5-41}$ \ergps; \citealt{Mouetal14, Mouetal15}) is consistent with the \ion{O}{8}/\ion{O}{7} line ratio as well as the enhanced past accretion rate in \sgrA \citep{Totani06}.   
    \item \textit{Tidal-Disruption driven}: Wind driven by a succession of Tidal Disruption Events (TDEs) taking place at the Galactic center (\citealt{Ko2020}; Piran \etal~\textit{in prep}.) could satisfy the required energy conditions provided that there was a period of $\sim 10$ Myr during which the Galactic TDE rate was slightly higher than expected (i.e. an event per $10^3$ years). With energy injection of a few $\times 10^{51}$ erg per event such an enhanced period of TDEs can supply the needed energy budget. The results of such wind  should resemble those of SN or accretion-driven wind discussed earlier. 
\end{itemize}
In all the above cases, irrespective of star-formation or AGN origin, we find that the jet/wind power has to be $\lesssim 10^{41}$ \ergps to produce symmetric FEBs and satisfy the \oviii/\ovii constraint. With such low power, the age of the Fermi Bubbles is estimated to be $\sim 10^{56}$ erg$/(10^{41} \mbox{\ergps}) \approx 30$ Myr. While a combined AGN and star-formation driven scenario looks like an attractive solution for the FEBs, we must note that the presence of one source can suppress the presence of the other \citep{Angles-Alcazar2021}. It is, therefore, more likely that either the star-formation or the AGN-driven wind/jet scenario (with limited power) is more natural.

\section*{Acknowledgements}
We thank Omar Bromberg, Karamveer Kaur, Dipanjan Mukherjee, Matteo Pais, and  Nicholas Stone for helpful discussions. KCS acknowledges support from the German Science Foundation via 
DFG/DIP grant STE 1869/2-1 GE 625/17-1 and Israeli Science Foundation(ISF) via grant No.2190/20. 
SM acknowledges Ramanujan Fellowship (No RJF/2020/000113) by SERB-DST, Govt. of India. SM also acknowledges the use of the computing resources made available by the Computer Centre (NOVA Cluster) of the Indian Institute of Astrophysics for this work. 
Some  simulations were carried out on PARAM Pravega cluster at IISc for which PS acknowledges a National Supercomputing Mission (NSM) grant from the Department of Science and Technology, India. PS also acknowledges a Swarnajayanti Fellowship (DST/SJF/PSA-03/2016-17). TP acknowledges support from Advanced ERC grants TReX and Multijets. 


\appendix 
\section{Grid structure in our simulations}
\label{app-sec:grid-structure}
\begin{figure}
    \centering
    \includegraphics[width=0.45\textwidth]{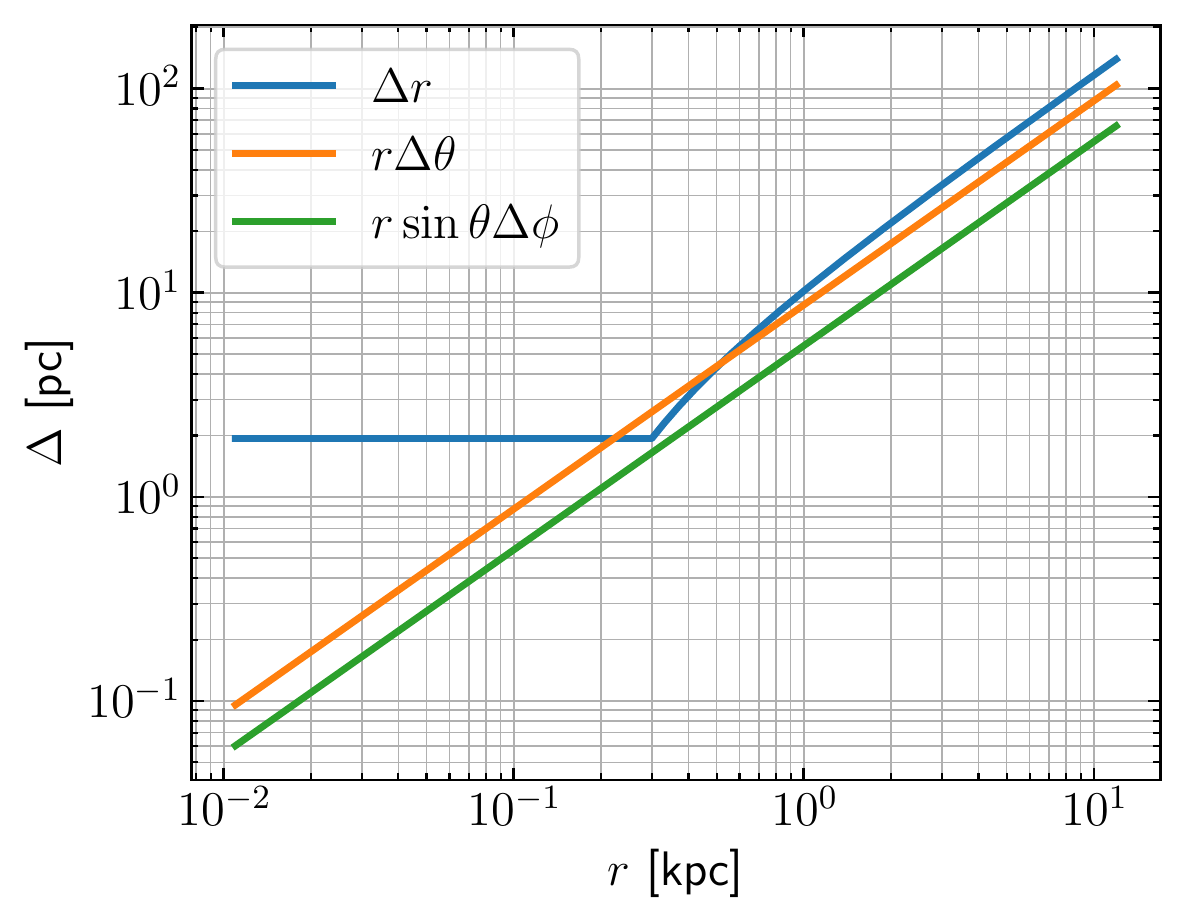}
    \caption{Resolutions in the $r$, $\theta$, and $\phi$-directions (here, $\theta=45^\circ$) in our spherical grid simulations. The resolution is $\Delta < 2$ pc at $r<200$ pc (in the ISM) and $\Delta \sim 0.2$ pc at $r=30$ pc (at the jet base).
    }
    \label{app-fig:grid}
\end{figure}

The grid structure in our spherical simulations is shown in figure \ref{app-fig:grid} along different directions. We resolve the central $200$ pc with a resolution $\Delta <2$ pc. The jet base ($r = 30$ pc) is resolved with $\Delta = 0.2$ pc, suitable for properly resolving the jet-collimation.
We ran a simulation in spherical coordinate with twice the spatial resolution
for the MP case ($L_j = 2\times 10^{41}$ \ergps) and found that the currently adopted resolution is suitable to produce a converged shape of the cocoons.

\section{Comparing with previous simulations}
\label{app-sec:prev-works}
In addition to the simulations in the spherical coordinates, we also perform simulations in 3D Cartesian coordinates. In this setup, the simulation box extends from $-9$ to $+9$ kpc in X- and Y- directions, and from $-11$ to $+11$ kpc in Z-direction. The ISM disk is set to lie on the X-Y plane. To understand the effects of resolution and compare them with other works, we run our MP case ($L_j = 2\times 10^{41}$ \ergps) at two resolutions. For the low resolution case, we use $\Delta_{\rm ISM} = 4$ pc resolution in the central $\pm 200$ pc in all the directions and $\Delta_{\rm CGM} = 50$ pc in the rest of the simulation box. In the high-resolution simulation, we increase the overall resolution by a factor of $2$ throughout the box. The jet was injected at an angle of $45^\circ$ from the galaxy rotation axis and in a region within a half opening angle of $5^\circ$ and $r\leq 30$ pc in the same fashion as in the main simulations (sec \ref{subsec:set-up}).

\begin{figure*}
    \centering
    \includegraphics[width=0.85\textwidth]{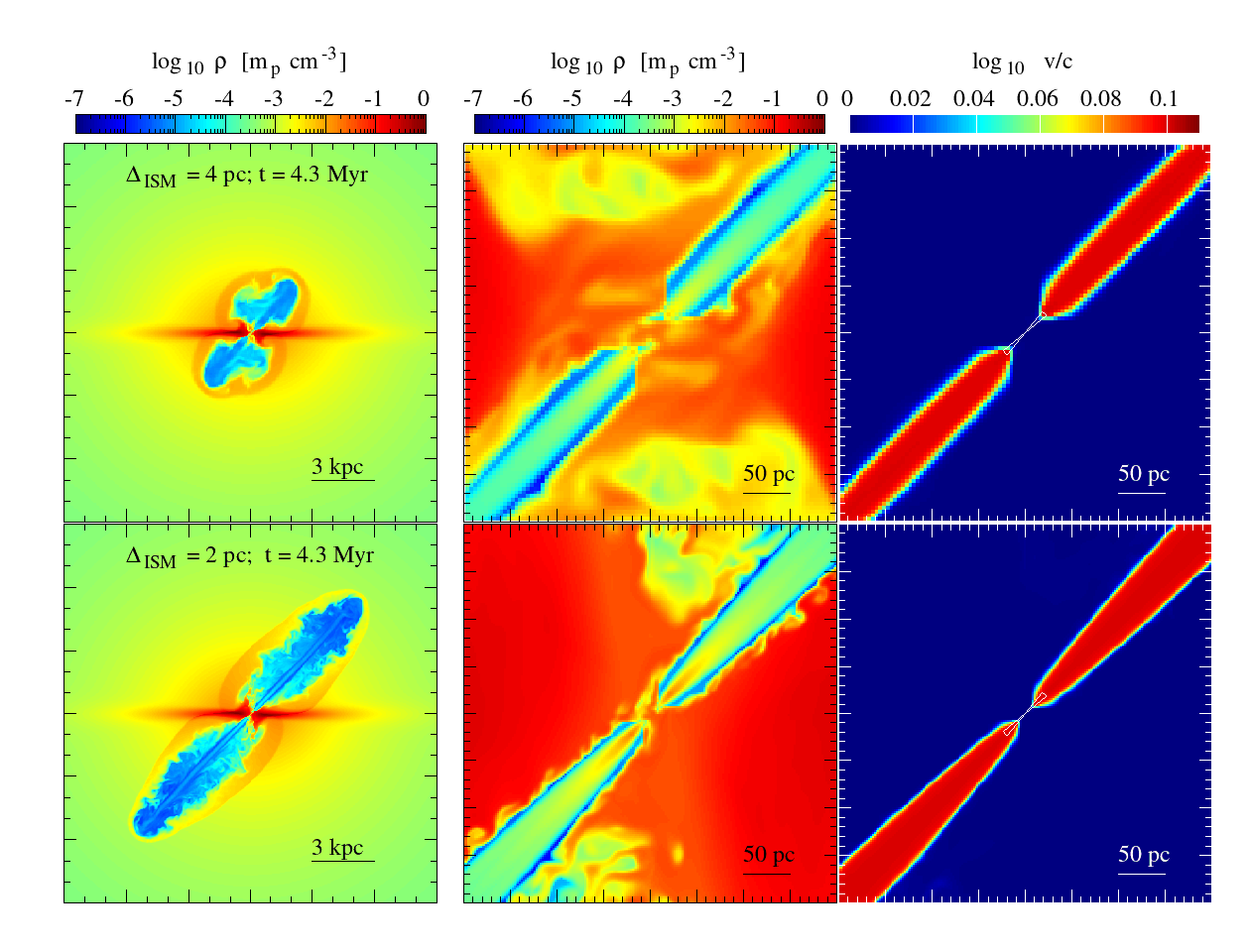}
    \caption{Comparing two resolutions for $L=2\times 10^{41}$ \ergps simulations. The simulations are run in Cartesian coordinates and the plots show a $X-Z$ slice at $y=\Delta/2$. The top panel shows results from the low-resolution run ($\Delta_{\rm ISM} = 4$ pc) and the bottom panel shows the results from the higher resolution run ($\Delta_{\rm ISM} = 2$ pc) at the same time, i.e. $t=4.3$ Myr. The first column shows the density map for the entire box, the second column shows a zoomed-in version of the density within the central $\pm 100$ pc, and the third column shows the zoomed-in velocity maps. The higher resolution jets are less dissipated owing to lower jet surface area and sharper jet-ISM boundary. The white cones represent the intended region of jet injection. }
    \label{app-fig:dens-resolution}
\end{figure*}

Figure \ref{app-fig:dens-resolution} shows the results for the two simulations. 
The velocity maps show that the jet widens as soon as it leaves the jet injection region. The widening is worse in the case of the lower-resolution simulation. The higher resolution simulation does a slightly better job at resolving the collimated structures at the jet base. 
The significant difference between the two runs suggests that much finer grid spacing is needed (for a Cartesian grid) to resolve the jet and in particular its collimation, which in turn determines its overall fate. 
The spherical coordinate system, on the other hand, naturally resolves the central base of the jet (see figure \ref{app-fig:grid}). 

Several simulations of jet-ISM interaction show dissipation of the jets in the ISM \citep[such as][]{Mukherjee2018,Tanner2022}. These simulations employ Cartesian coordinates with a resolution of $\sim 5-10$ pc to resolve the jet with $\sim 30$ pc wide base. 
As we see in figure \ref{app-fig:dens-resolution}, comparable simulations with grid spacing of $2$ pc and $4$ pc 
did not converge, with the higher resolution run showing more collimation. Based on these simulations, we speculate that the dissipation seen in the earlier jet-ISM simulations could be partially due to an insufficient spatial resolution.

There can be other reasons for the difference between our results and the results  of previous simulations. The previous simulations launched wide angle ($\sim 20^\circ$) outflows. Such a high opening angle for the outflow increases the jet-ISM interaction. However, this is  insufficient by itself to lead to choking (see Eq. 7 and Fig. \ref{fig:therm-condition}). 

Another important aspect that separates the previous simulations from ours is the clumpiness in the ISM. The ISM itself was assumed to be extended till $\sim 0.5-1$ kpc and  significantly clumpy (volume filling factor $\sim 0.2-0.5$) with clump size as large as $\sim 300$ pc. 
However, the volume filling fraction for the clumpy  ISM gas, $\sim 10^{-3}$ \citep{Draine2011}, is much lower in the Milky Way. 
\section{e-p non-equilibrium}
\label{app-sec:ep-noneq}
We solve for $T_e$ and $T_p$ considering only Coulomb interaction between electrons and protons \citep{Braginskii1965} i.e.,
\begin{eqnarray}
    \frac{3}{2} n_e \frac{dT_e}{dt} &=& - 3\frac{m_e}{m_p} \frac{n}{\tau_e} \left(T_e-T_i\right) \nonumber \\
    \frac{3}{2} n_i \frac{dT_i}{dt} &=&  3\frac{m_e}{m_p} \frac{n}{\tau_e} \left(T_e-T_i\right) 
\end{eqnarray}
where, $\tau_e = 3.44\times 10^5 \frac{(k_B T_e/\mbox{eV})^{3/2}}{n \lambda}$, $\lambda \approx 15$, $n = \rho/(0.6\: m_p)$ is the particle number density, $n_e = \rho/(1.15\: m_p)$ is the electron number density, and $n_i = \rho/(1.27\: m_p)$ is the ion number density for a fully ionized plasma. These equations can be solved for a given set of initial conditions.
The initial values for $T_e$ and $T_p$ are assumed to be the temperatures in a post-shocked gas and are calculated following \cite{Vink2015} (their Eq 19 and 21, which are based on adiabatic heating and thermalization of electrons at shocks and match observations of collisionless shocks) for a given Mach number of the shock. The evolution of the temperatures has been shown in the left panel of Fig \ref{fig:e-p-noneq}. Different  colors show different starting values for the post-shock gas ($t_{\rm dyn}$ is the dynamical time for $r_s = 10$ kpc). The figure shows that at early times ($t=t_{\rm dyn}/3 \approx 4$ Myr) of the blast wave, the starting $T_e/T_p \approx 0.3$. The temperatures, however, quickly become equal within $\sim 1$ Myr. Incidentally, the equilibrium time scale at any stage of the shock is about the same, for the given density profile. This implies that for older shocks (age $\gg 1$ Myr), the electrons and protons will have enough time to be in equilibrium. Given that the dynamical time of the FEBs is $\sim 12$ Myr for the EP model, we expect $T_e=T_p= T_{\rm eq}$ in FEBs.  

\bibliography{fb3d}
\bibliographystyle{aasjournal}

\end{document}